\documentclass{PoS}

\usepackage{subfig}

\title{Top quark physics at CDF}

\ShortTitle{Top quark physics at CDF}

\author{\speaker{Karolos Potamianos}%
       \thanks{On behalf of the CDFcollaboration}\\
       Purdue University\\
       Fermilab, Batavia, IL, USA\\
       E-mail: \email{karolos.potamianos@cern.ch}}

\abstract{
We present the recent results of top-quark physics using up to 6~fb$^{-1}$ of $p\bar{p}$ collisions analyzed by the CDF collaboration. The large number of top quark events analyzed, of the order of several thousands, allows stringent checks of the standard model predictions. Also, the top quark is widely believed to be a window to new physics. We present the latest measurements of top quark intrinsic properties as well as direct searches for new physics in the top sector.}

\FullConference{The 2011 Europhysics Conference on High Energy Physics-HEP 2011,\\
		July 21-27, 2011\\
		Grenoble, Rh\^one-Alpes France}

\newcommand{\gevcc}{\mbox{GeV/$c^2$}}
\newcommand{\ttbar}{\mbox{$t\bar{t}$}}
\newcommand{\ppbar}{\mbox{$p\bar{p}$}}
\newcommand{\met}{\mbox{$\protect \raisebox{0.3ex}{$\not$}E_T$}}
\newcommand{\mpt}{\mbox{$\protect \raisebox{0.3ex}{$\not$}P_T^{\rm tr}$}}
\newcommand{\invfb}{\mbox{fb$^{-1}$}}

\begin{document}

\section{Introduction}

The top quark is the third generation up-type quark. It was discovered in 1995 by the CDF and D\O\ experiments at the Tevatron~\cite{TopQuarkObs}, in events where it is produced together with an antitop quark. It has a charge of $+\frac{2}{3}e$. With a mass of \mbox{$173.2\pm 0.9~\gevcc$}~\cite{Lancaster:2011wr}, it is the most massive elementary particle known to date. 
According to the standard model (SM), it has a Yukawa coupling to the Higgs boson close to unity, which hints to a possible special role in electroweak symmetry breaking. 
It interacts primarily by the strong interaction but also through the weak force.
In 2009, the CDF and D\O\ experiments observed the production of a single top quark through the weak interaction~\cite{SingleTopObs}.

Since its observation, the top quark has been extensively studied, and is currently one of the main physics programs at the Tevatron. It has a very short lifetime ($\sim 10^{-25}$~s) and hence decays before hadronizing, providing a unique opportunity to study a bare quark. Therefore, top quark physics is a window into new physics. We present the latest results from the CDF collaboration.

\section{Top Quark Physics}

At the Tevatron \ppbar\ collider, the top quark is mainly produced in top-antitop pairs (\ttbar), through quark-antiquark (85\% contribution to $\ttbar$) and gluon-gluon (15\% contribution) fusion.
The top quark decays through the weak force almost exclusively into a $W$ boson and a $b$ quark.

The production of top pairs is studied according to the $\ttbar$ decay mode: dilepton, lepton+jets and all-hadronic, depending on whether two, one, or none of the $W$ decayed leptonically. The lepton+jets is considered the golden channel because it combined a good branching fraction ($\sim30\%$) with a good signal to background ratio. The dilepton ($ee,e\mu,\mu\mu$) channel is the cleanest but suffers from a small branching fraction ($\sim5\%$). The all-hadronic channel has the largest branching fraction ($\sim45\%$) but suffers from a huge level of background from QCD multijet production.
An additional channel, \met+jets, is used to recover events where the $e$ or $\mu$ is not identified in the detector, or when the $\tau$ decays hadronically; it also suffers from a large QCD background.

\section{Top Quark Production}

The SM allows predicting the \ttbar\ production cross-section $\sigma(\ttbar)$. Measurements of this cross-section test QCD calculations and allow probing new physics beyond the standard model, since new physics can enhance \ttbar\ production.
Next to leading order (NLO) predictions estimate the \ttbar\ cross-section to be $\sigma(\ttbar) = 7.45^{+0.72}_{-0.63}$~pb for a top quark mass of $172.5~\gevcc$~\cite{Moch:2008ai}.

\begin{figure}[t]
\centering
\subfloat[\label{fig:ttbar_xsec}]{\includegraphics[width=.5\linewidth]{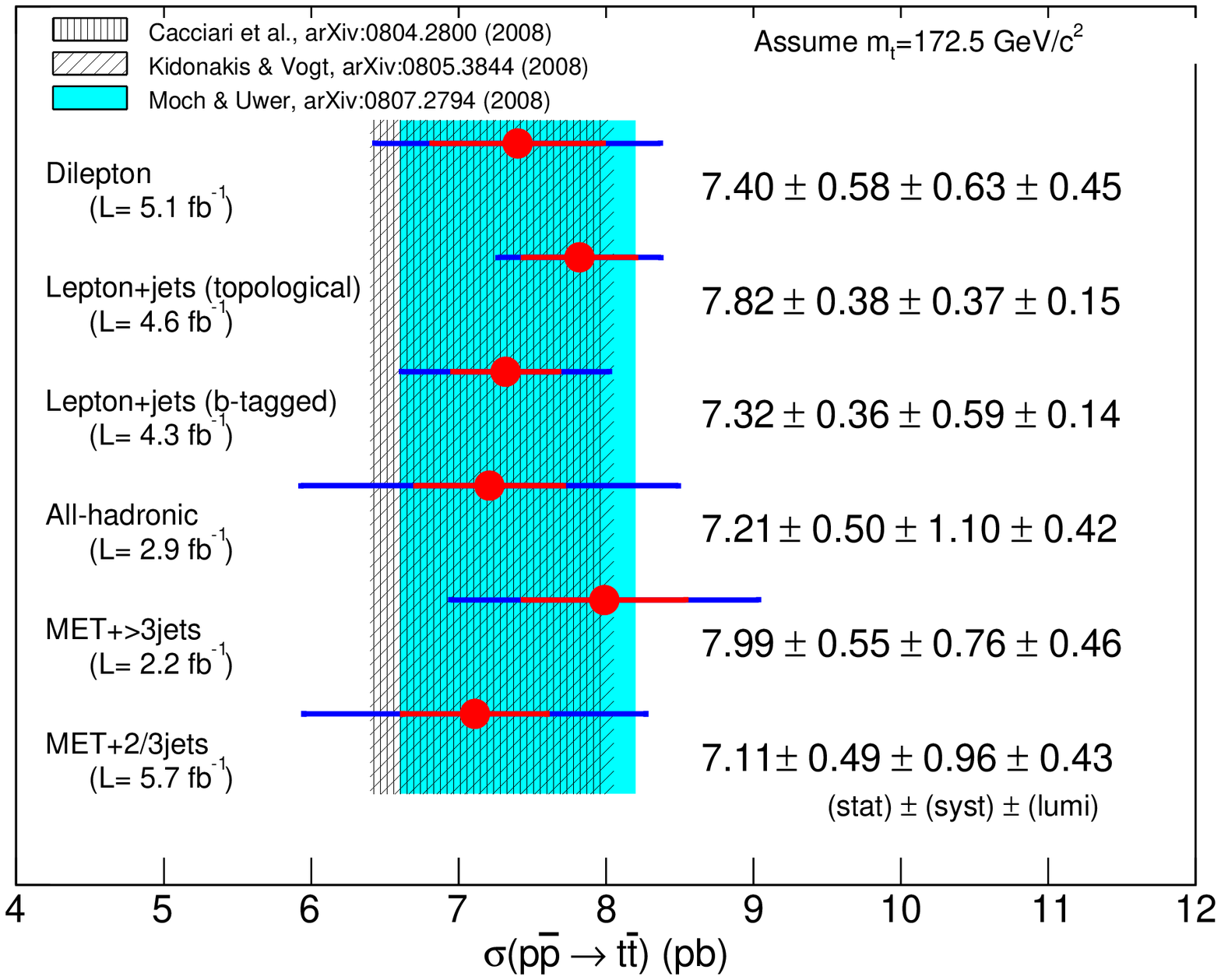}}\hfill
\subfloat[\label{fig:top_mass}]{\includegraphics[width=.3\linewidth]{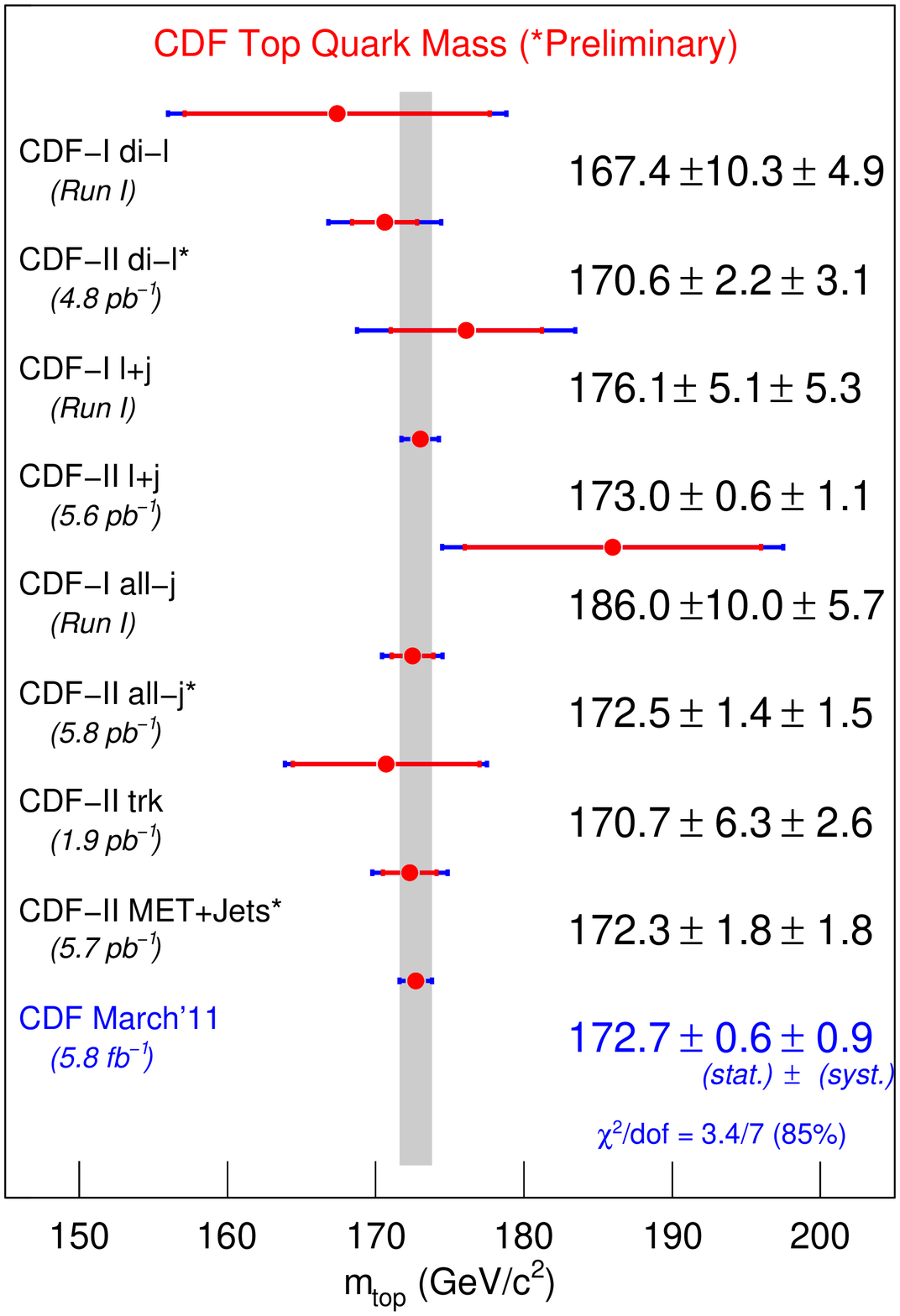}}
\caption{Summary of CDF measurements of $\sigma(\ttbar)$ (a) and of $m_t$ (b).}
\end{figure}

The most precise measurements of $\sigma(\ttbar)$ come from the lepton+jets signature~\cite{Aaltonen:2010ic}. 
This analysis uses two complementary methods: a $b$-jet tagging measurement and a topological approach based on a neural network.
The combined result yields $\sigma(\ttbar) = 7.70 \pm 0.52$~pb for a top-quark mass of $172.5~\gevcc$.
In the dilepton channel, we measure a cross-section of $\sigma(\ttbar) = 7.4 \pm 1.0$~pb~\cite{cdf10163} using $5.1~\invfb$ of data. After requiring that at least one of the jets originating from a $b$ quark by identified by the {\sc SecVtx} $b$-tagging algorithm~\cite{PhysRevD.71.052003}, we measure $\sigma(\ttbar) = 7.3 \pm 0.9$~pb. In the all-hadronic channel, we measure a cross-section of $\sigma(\ttbar) = 7.2 \pm 1.2$~pb~\cite{Aaltonen:2010pe}. Two results in the \met+jets signature yield $\sigma(\ttbar) = 7.1 \pm 1.2$~pb (2/3 jets in $5.7~\invfb$)~\cite{cdf10237} and $\sigma(\ttbar) = 8.0 \pm 1.0$~pb ($>3$ jets in $2.2~\invfb$)~\cite{Aaltonen:2011tm}.
In the hadronic $\tau$+jets channel, we obtain using $2.2~\invfb$ of data $\sigma(\ttbar) = 8.8 \pm 4.0$~pb~\cite{cdf10562}. Figure~\ref{fig:ttbar_xsec} summarizes most of the cross-section measurements at CDF.

\section{Top Quark Properties}

\begin{figure}[b]
\centering
\begin{tabular}{ccc}
\subfloat[]{\includegraphics[width=.4\linewidth]{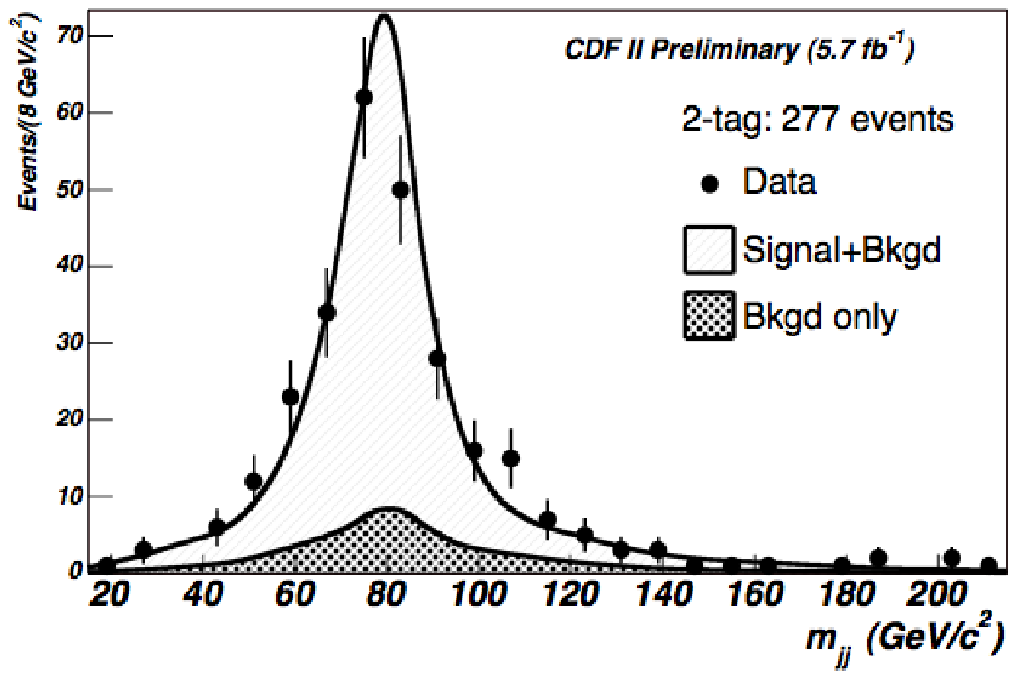}} & \hspace{1cm} &
\subfloat[]{\includegraphics[width=.3\linewidth]{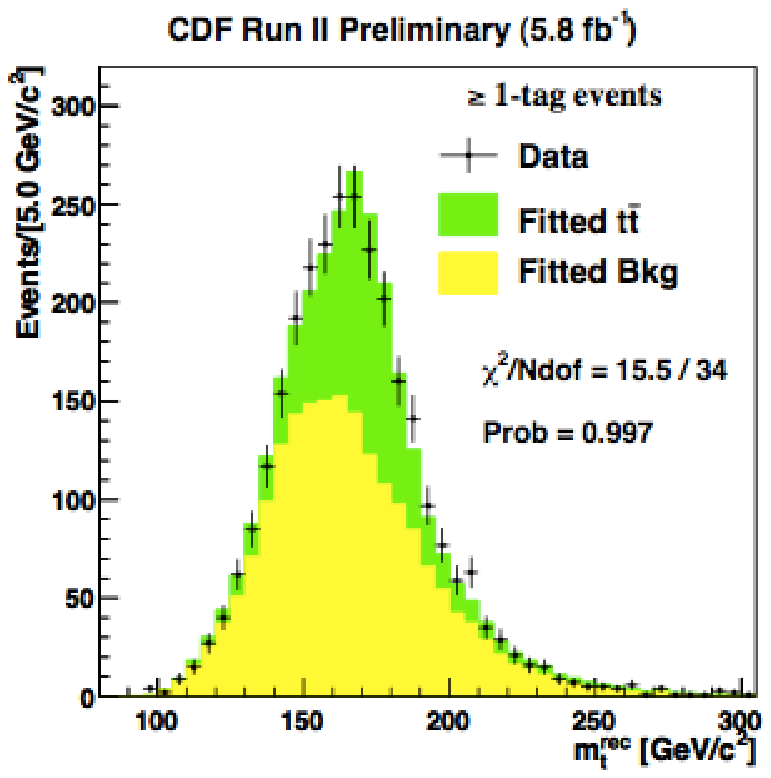}} \\
\end{tabular}
\caption{\label{fig:mtopNew}Top mass measurements in the \met+jets (a) and the all-hadronic signatures (b).}
\end{figure}

\begin{figure}[b]
\centering
\begin{tabular}{ccc}
\subfloat[]{\includegraphics[width=.4\linewidth]{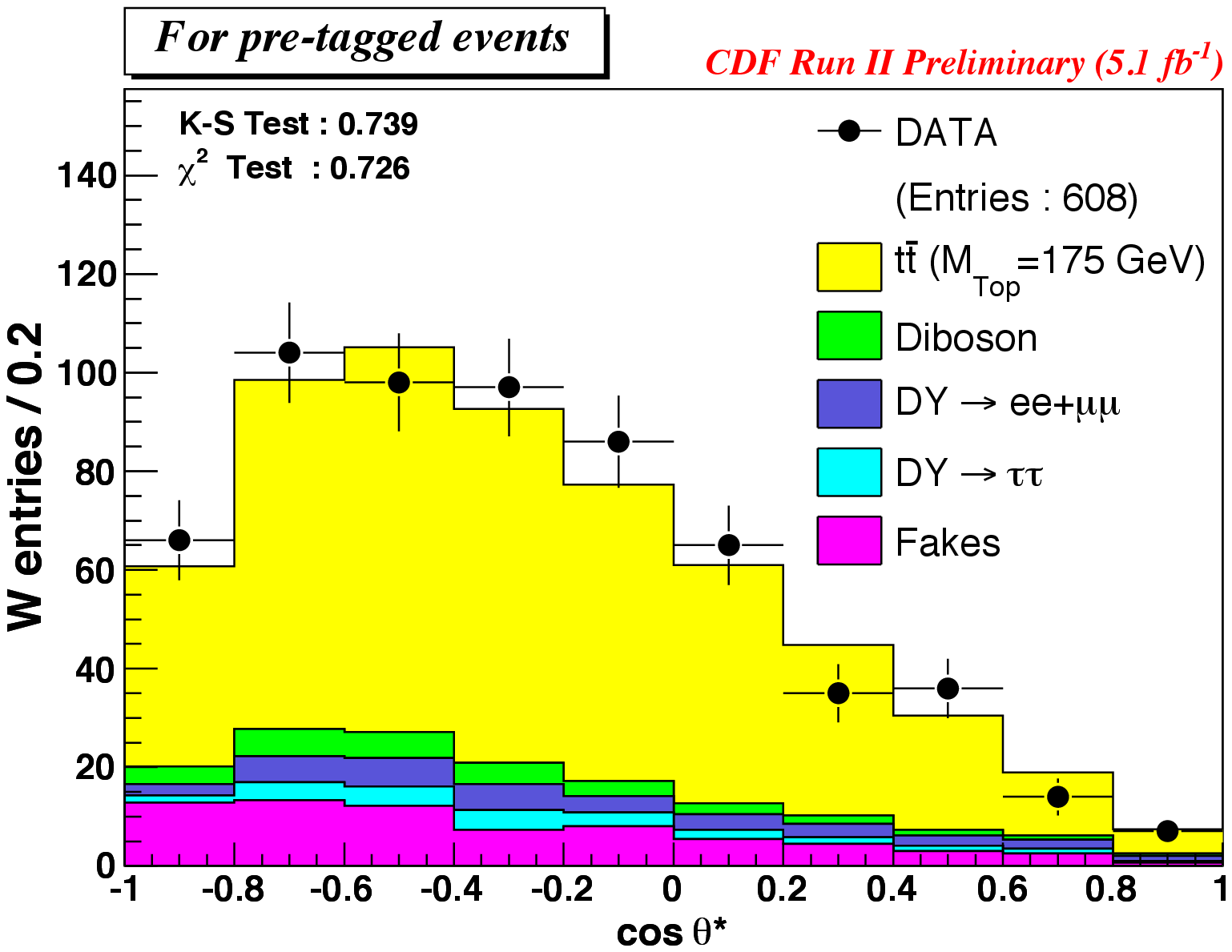}} & \hspace{1cm} &
\subfloat[]{\includegraphics[width=.4\linewidth]{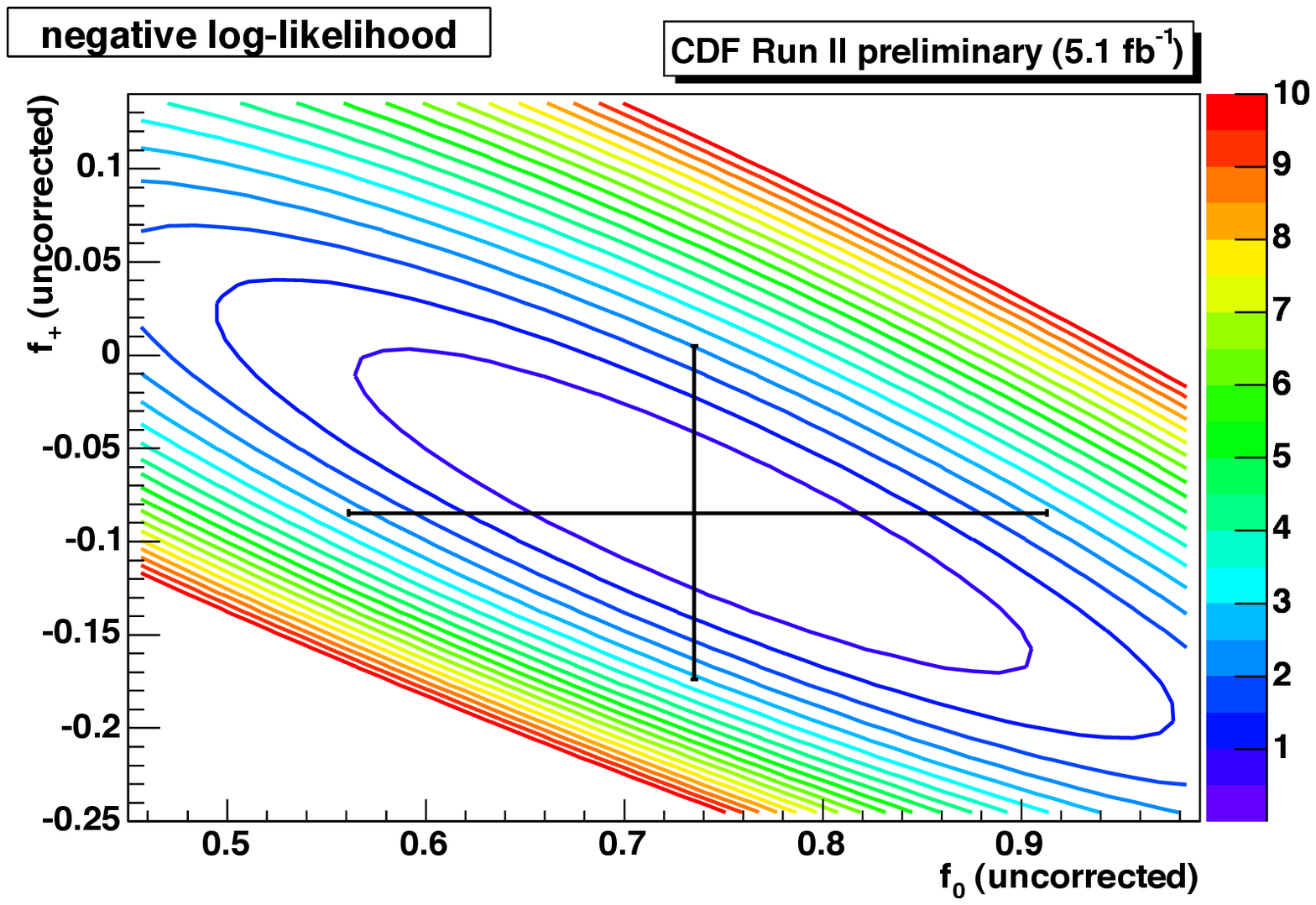}}\\
\subfloat[]{\includegraphics[width=.3\linewidth]{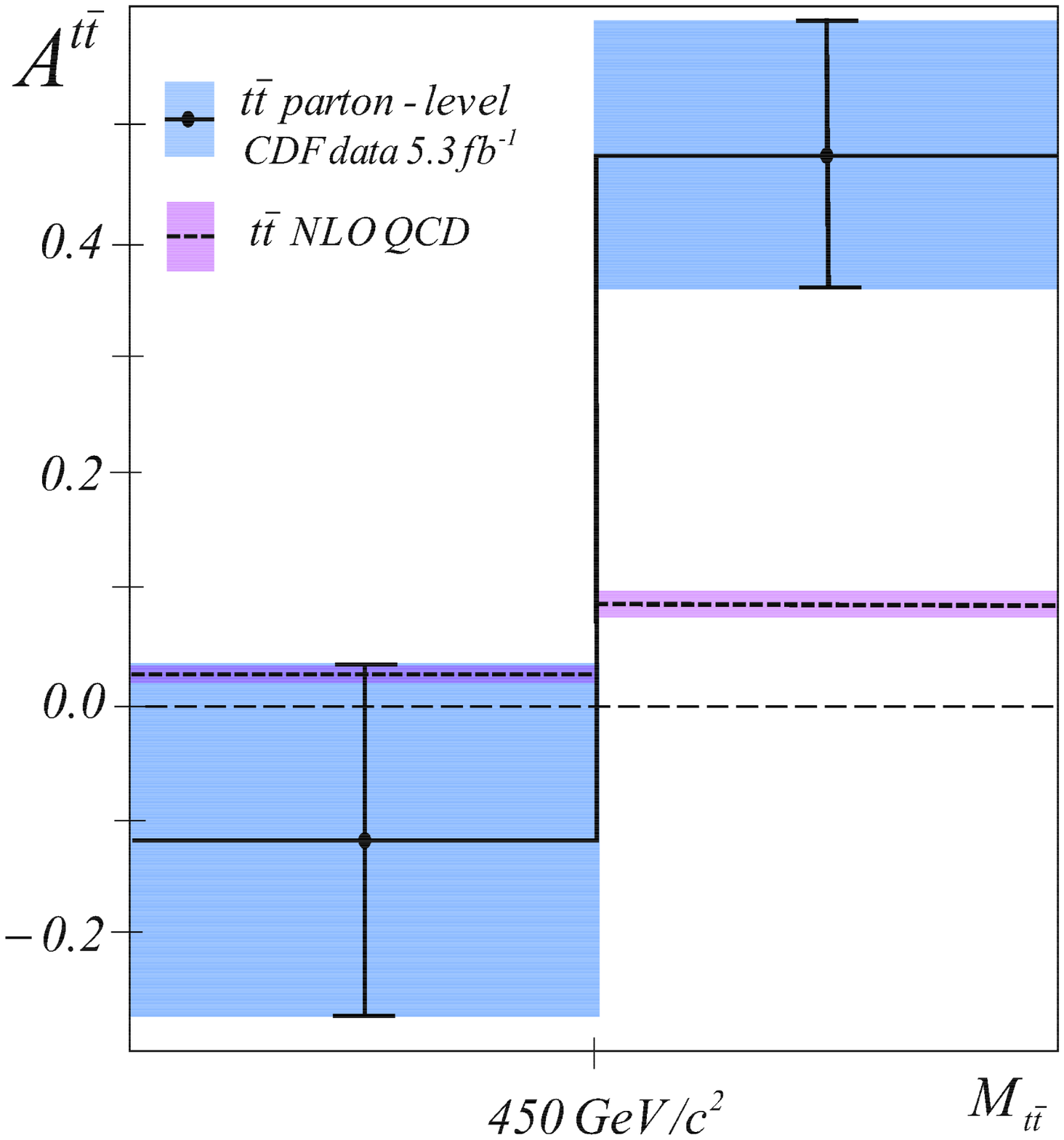}} & &
\subfloat[]{\includegraphics[width=.5\linewidth]{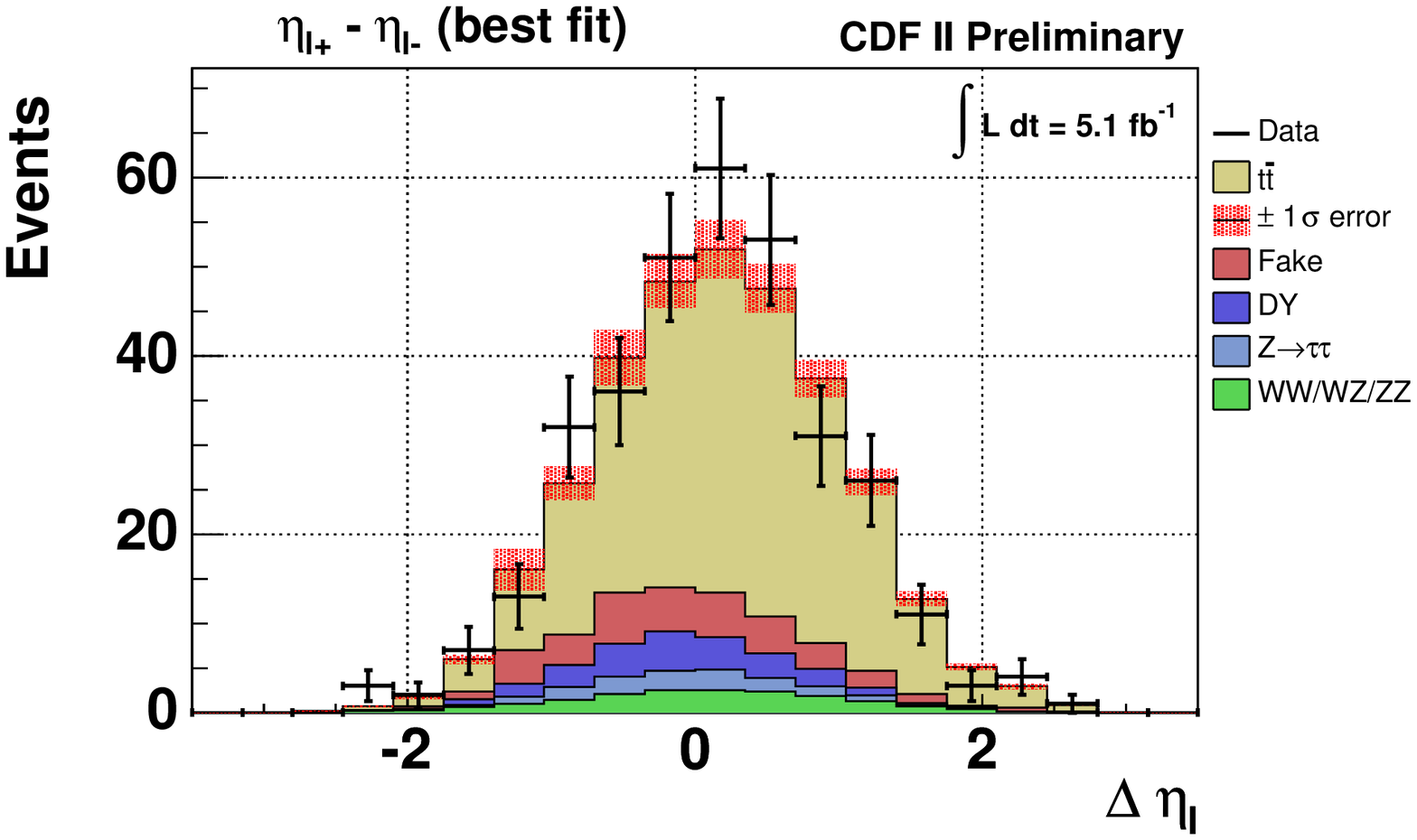}}
\end{tabular}
\caption{\label{fig:Whelicity}\label{fig:Afb} Top quark properties measurements at CDF: $W$ boson polarization in the dilepton channel (a,b) and forward-backward asymmetry in the lepton+jets (c) and dilepton (d) channels.}
\end{figure}

Among the properties of the top quark, its mass is the most studied. It is a fundamental parameter of the SM because it constrains the mass of the Higgs boson. Recent CDF results include a measurement in the \met+jets and all hadronic final states using $5.7~\invfb$, with a top mass respectively of $m_t = 172.3 \pm 2.6~\gevcc$~\cite{Aaltonen:2011kx} and  $m_t = 172.5 \pm 2.1~\gevcc$~\cite{cdf10456}. These results, shown in figure~\ref{fig:mtopNew}, are combined with past determinations of the top mass and yield $m_t = 172.7 \pm 1.1~\gevcc$~\cite{cdf10444}, reaching a precision of $\Delta m_t/m_t = 0.63\%$ and a uncertainty on the top quark of $1.1~\gevcc$. Figure~\ref{fig:top_mass} summarizes these results.
The CDF collaboration also reported the first top mass measurement in the $\tau$+jets channel using $2.2~\invfb$ of data, yielding $m_t = 172.7 \pm 10~\gevcc$~\cite{cdf10562}.
The mass difference between the top and antitop quark is measured to be $m_t - m_{\bar{t}} = -3.3 \pm 1.7~\gevcc$~\cite{Aaltonen:2011wr}, consistent with CPT symmetry.

Additional properties studied at CDF include the $W$ boson polarization of the top quark decay and the forward-backward asymmetry. The $V-A$ structure of the weak interaction predicts that the $W^+$ from $t\to W^+b$ decays are either longitudinally polarized ($\sim70\%$) or left-handed ($\sim30\%$), while right-handed $W$ bosons are heavily suppressed (and forbidden in the limit of a massless $b$~quark). The latest CDF measurements in the dilepton channel yield a longitudinal fraction \mbox{$f_0 = 0.74^{+0.19}_{-0.18}$} and a right-handed fraction of $f_+ = -0.09 \pm 0.1$. 

The forward-backward asymmetry of top pair production $A_{fb}$ is due to NLO effects and is predicted to be about $8\%$~\cite{Antunano:2007da}. In the lepton+jets channel, the corrected parton level asymmetry is $16\pm8\%$ ($48\pm11\%$ for events with $m_{\ttbar}>450~\gevcc$)~\cite{Aaltonen:2011kc}, while an asymmetry of $42\pm16\%$ is measured in the dilepton channel~\cite{cdf10436}. 
The combination of these results yields $A_{fb}= 20 \pm 7\%$~\cite{cdf10584}.

\section{Searches for New Physics}

\begin{figure}
\includegraphics[height=3.5cm]{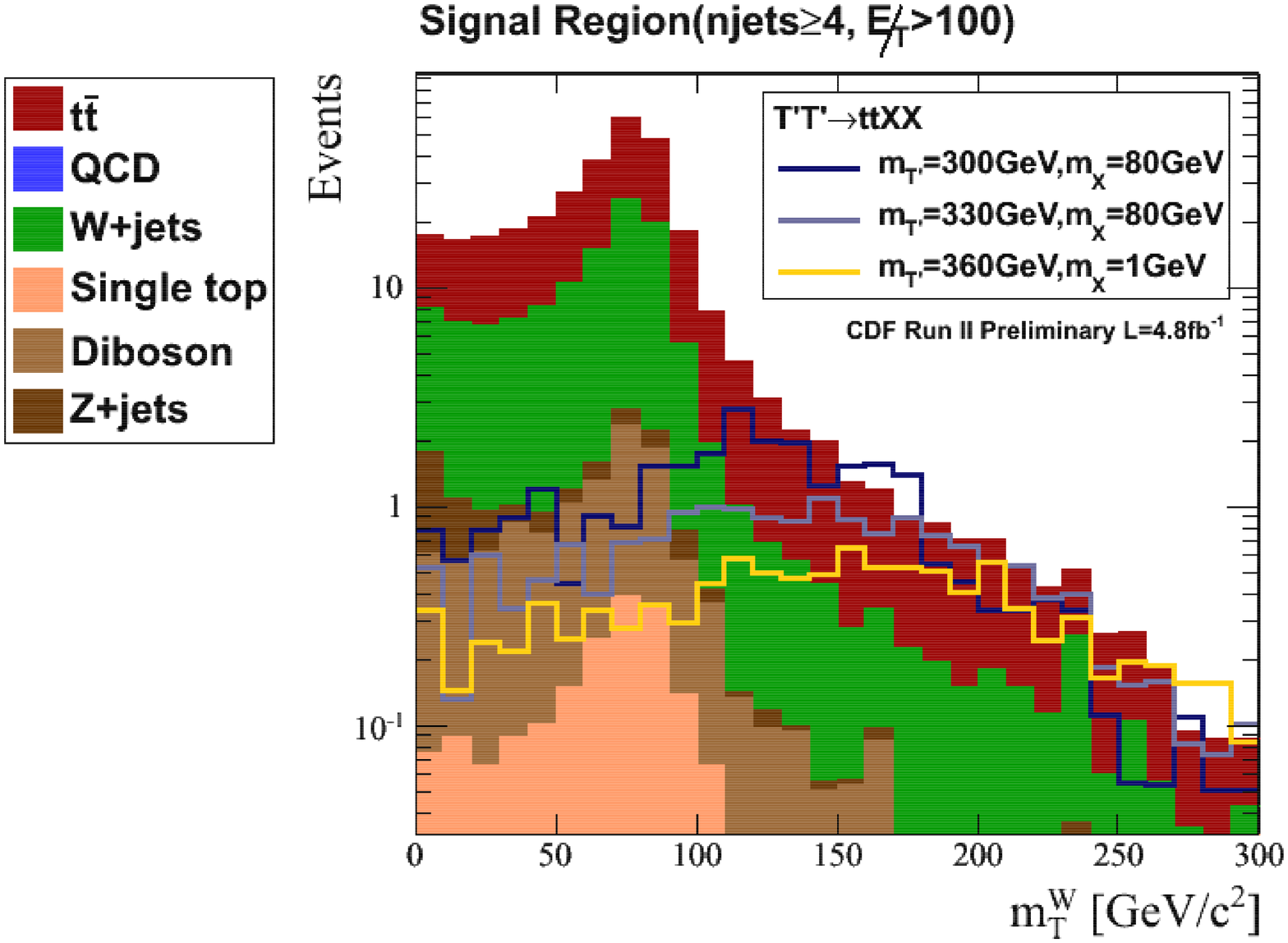}\hfill\includegraphics[height=3.5cm]{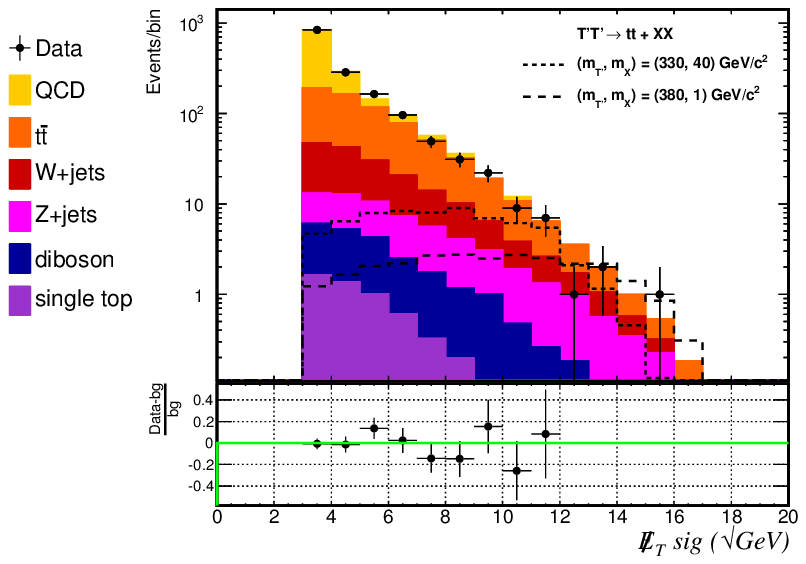}\hfill\includegraphics[height=3.5cm]{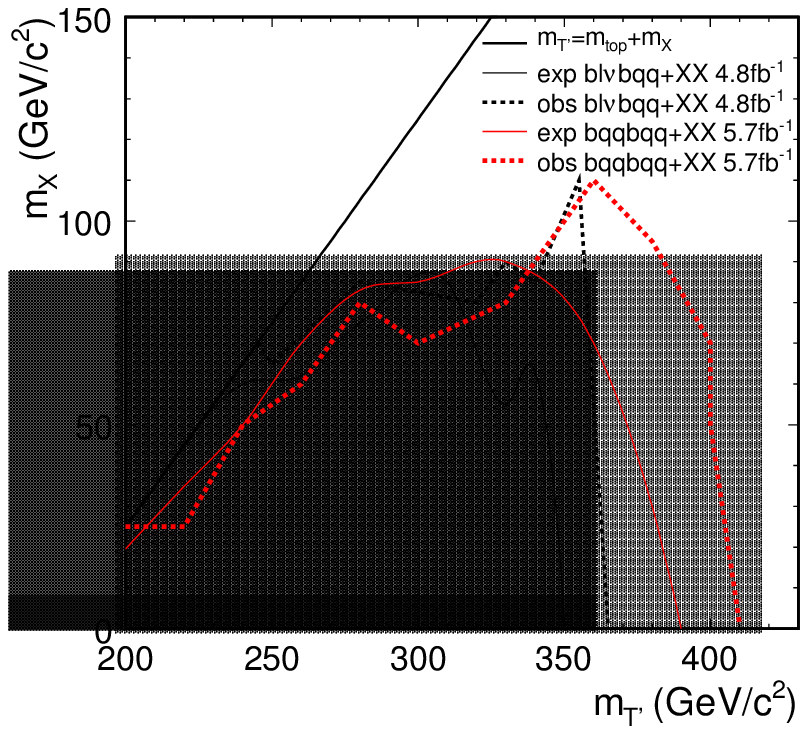}
\caption{\label{fig:tprime}Searches for fourth generation quarks $t^\prime$ decaying via $t^\prime \to t + X$ where $X$ is dark matter in the lepton+jets (left) and \met+jets (center) signature. The plot on the right shows the results of these searches. }
\end{figure}

Several searches look for new physics in the top sector. CDF has searched for a heavy top quark $t^\prime\to Wb$ in the lepton+jets signature~\cite{cdf10395}. Looking at $5.6~\invfb$ of CDF data, we exclude the SM fourth generation $t^\prime$ quark with a mass below $358~\gevcc$ at 95\% confidence level.
Another result searches for dark matter through the production of fourth generation quarks $t^\prime$ decaying via $t^\prime \to t + X$ where $X$ is dark matter~\cite{Aaltonen:2011rr}. Analyzing 4.8~\invfb\ of CDF data, we exclude \mbox{$m_{t^\prime} < 360~\gevcc$} at 95\% confidence level for $m_X  < 100~\gevcc$. 
The same search is done in the \met+jets channel, thus studying events where both $W$ bosons decay hadronically~\cite{Aaltonen:2011na}. To distinguish events in which the \met\ is due to the dark matter form those in which it is due to the mis-measurement of the jet energies, the latter uses the missing momentum flow (\mpt) in the tracker as a complementary tool to the \met\ in the calorimeter. Looking at $5.7~\invfb$, we exclude \mbox{$m_{t^\prime} < 400~\gevcc$} at 95\% confidence level for $m_X  < 70~\gevcc$. 
Figure~\ref{fig:tprime} shows these results.

\section{Conclusion}

We have presented the latest measurements by the CDF collaboration of top quark intrinsic properties as well as direct searches for new physics in the top sector.
The large number of top quark events analyzed, of the order of several thousands, allows stringent checks of the standard model predictions. 
Many of these results will be updated with the final CDF dataset.


\begin{thebibliography}{99}
\bibitem{TopQuarkObs}
S.~Abachi {\it {\it et~al.},}, \newblock {Observation of the top quark}, \newblock {\em Phys. Rev. Lett.}, 74:2632--2637, 1995;
F.~Abe {\it {\it et~al.},}, \newblock {Observation of top quark production in $\bar{p}p$ collisions}, \newblock {\em Phys. Rev. Lett.}, 74:2626--2631, 1995.

\bibitem{Lancaster:2011wr}
{Tevatron Electroweak Working Group},
\newblock {Combination of CDF and D\O\ results on the mass of the top quark
  using up to 5.8~fb$^{-1}$ of data},
\newblock 2011.

\bibitem{SingleTopObs}
T.~Aaltonen {\it et~al.},
\newblock {First Observation of Electroweak Single Top Quark Production},
\newblock {\em Phys.Rev.Lett.}, 103:092002, 2009; 
V.~M. Abazov {\it et~al.},
\newblock {Observation of Single Top Quark Production},
\newblock 2009.

\bibitem{Moch:2008ai}
S.~Moch and P.~Uwer.
\newblock {Heavy-quark pair production at two loops in QCD},
\newblock {\em Nucl.Phys.Proc.Suppl.}, 183:75--80, 2008.

\bibitem{Aaltonen:2010ic}
T.~Aaltonen {\it et~al.},
\newblock {First Measurement of the Ratio $\sigma_(t\bar{t}) /
  \sigma_(Z/\gamma**\to \ell\ell)$ and Precise Extraction of the $t\bar{t}$
  Cross Section},
\newblock {\em Phys.Rev.Lett.}, 105:012001, 2010.

\bibitem{cdf10163}
T.~Aaltonen {\it et~al.},
\newblock {Top Dilepton Cross Section in $5.1~\invfb$ using the DIL Selection},
\newblock CDF/PUB/TOP/PUBLIC/10163, 2011.

\bibitem{PhysRevD.71.052003}
D.~Acosta {\it et~al.},
\newblock {Measurement of the $t\bar{t}$ production cross section in $p\bar{p}$
  collisions at $\sqrt{s} = 1.96$ TeV using lepton + jets events with secondary
  vertex $b$-tagging},
\newblock {\em Phys. Rev. D}, 71(5):052003, 2005.

\bibitem{Aaltonen:2010pe}
T.~Aaltonen {\it et~al.},
\newblock {Measurement of the Top Quark Mass and $p\bar{p} \to t\bar{t}$ Cross
  Section in the All-Hadronic Mode with the CDFII Detector},
\newblock {\em Phys.Rev.}, D81:052011, 2010.

\bibitem{cdf10237}
T.~Aaltonen {\it et~al.},
\newblock {Measuring the Top Pair Background to the Higgs Search in the
  MET+b-jets Channel with L=$5.7~\invfb$},
\newblock CDF/PUB/TOP/PUBLIC/10237, 2011.

\bibitem{Aaltonen:2011tm}
T.~Aaltonen {\it et~al.},
\newblock {Measurement of the $t\bar{t}$ production cross section in $p\bar p$
  collisions at $\sqrt{s}=1.96$~TeV using events with large Missing $E_T$ and
  jets},
\newblock {\em Phys.Rev.}, D84:032003, 2011.

\bibitem{cdf10562}
T.~Aaltonen {\it et~al.},
\newblock {Measurements of Top Quark Properties in the $\tau$ + jets decay
  channel at CDF},
\newblock CDF/PUB/TOP/PUBLIC/10562, 2011.

\bibitem{Aaltonen:2011kx}
T.~Aaltonen {\it et~al.},
\newblock {Top-quark mass measurement using events with missing transverse
  energy and jets at CDF},
\newblock 2011.

\bibitem{cdf10456}
T.~Aaltonen {\it et~al.},
\newblock {Measurement of the top quark mass with in situ jet energy scale
  calibration in the all-hadronic channel using the Template Method with
  $5.8~\invfb$},
\newblock CDF/PUB/TOP/PUBLIC/10456, 2011.

\bibitem{cdf10444}
T.~Aaltonen {\it et~al.},
\newblock {Combination of CDF top quark mass measurements},
\newblock CDF/PUB/TOP/PUBLIC/10444, 2011.

\bibitem{Aaltonen:2011wr}
T.~Aaltonen {\it et~al.},
\newblock {Measurement of the mass difference between $t$ and $\bar{t}$
  quarks},
\newblock {\em Phys.Rev.Lett.}, 106:152001, 2011.

\bibitem{Antunano:2007da}
Oscar Antunano, Johann~H. Kuhn, and German Rodrigo.
\newblock {Top quarks, axigluons and charge asymmetries at hadron colliders},
\newblock {\em Phys.Rev.}, D77:014003, 2008.



\bibitem{Aaltonen:2011kc}
T.~Aaltonen {\it et~al},
\newblock {Evidence for a Mass Dependent Forward-Backward Asymmetry in Top
  Quark Pair Production},
\newblock {\em Phys.Rev.}, D83:112003, 2011.

\bibitem{cdf10436}
T.~Aaltonen {\it et~al.},
\newblock {Measurement of the Forward Backward Asymmetry in Top Pair Production
  in the Dilepton Decay Channel using $5.1~\invfb$},
\newblock CDF/PUB/TOP/PUBLIC/10436, 2011.

\bibitem{cdf10584}
T.~Aaltonen {\it et~al.},
\newblock {Combination of the Forward-Backward Asymmetry in the Top Pair
  Production from L+J and DIL Channels using $5~\invfb$},
\newblock CDF/PUB/TOP/PUBLIC/10584, 2011.

\bibitem{cdf10395}
T.~Aaltonen {\it et~al.},
\newblock {Search for Heavy Top $t^\prime \to Wq$ in Lepton Plus Jets Events in
  $\int\mathcal{L}dt = 5.6~\invfb$},
\newblock CDF/PUB/TOP/PUBLIC/10395, 2011.

\bibitem{Aaltonen:2011rr}
T.~Aaltonen {\it et~al.},
\newblock {Search for Production of Heavy Particles Decaying to Top Quarks and
  Invisible Particles in $p\bar{p}$ collisions at $\sqrt{s}=1.96$ TeV},
\newblock {\em Phys.Rev.Lett.}, 106:191801, 2011.

\bibitem{Aaltonen:2011na}
T.~Aaltonen {\it et~al.},
\newblock {Search for New $T^\prime$ Particles in Final States with Large Jet
  Multiplicities and Missing Transverse Energy in ppbar Collisions at sqrt(s) =
  1.96 TeV},
\newblock {\em Phys.Rev.Lett.}, 107:191803, 2011.

\end{thebibliography}
\end{document}